\documentclass[preprint,showpacs,preprintnumbers,amsmath,amssymb]{revtex4}

\usepackage[section]{placeins}
\usepackage{graphicx}
\usepackage{dcolumn}
\usepackage{bm}

\usepackage{epsfig}
\usepackage{amssymb}
\usepackage{amsmath}
\usepackage{flafter}
\usepackage{array}

\usepackage[dvips,usenames]{color}

\newlength{\dinwidth}
\newlength{\dinmargin}
\begin{document}
\title{Large Strong Phases and CP Violation  in the Annihilation
Processes $\bar{B}^0\rightarrow K^+K^-$, $K^{*\pm}K^\mp$,
$K^{*+}K^{*-}$}

\author{Fang Su$^{a,b}$,  Yue-Liang Wu$^a$,
 Ya-Dong Yang$^c$, and Ci Zhuang$^{a,b}$} \affiliation{$^a$ Institute of Theoretical Physics,
 Chinese Academy of Science,
 Beijing 100080,  China\\
$^b$ Graduate School of the Chinese Academy of Science, Beijing,
100039, P.~R. China\\
 $^c$ Department
of Physics, Henan Normal University,
Xinxiang, Henan 453007, China\\
}

\begin{abstract}
\noindent The strong phases and CP violation in the rare
$\bar{B}^0\rightarrow K^+K^-$, $K^{*\pm}K^\mp$, $K^{*+}K^{*-}$
decays are investigated. As these decays proceed only via
annihilation type diagrams in the standard model~(SM), a dynamical
gluon mass is introduced to avoid the infrared divergence in the
soft endpoint regions. The Cutkosky rule is adopted to deal with a
physical-region singularity of the on mass-shell quark
propagators, which leads to a big imaginary part and hence a large
strong phase. As a consequence, large CP asymmetries are predicted
in those decay modes due to a big interference between the
annihilation amplitudes from penguin and tree operators, which may
be tested in future more precise experiments.
\end{abstract}

 \maketitle

\section{Introduction}

Charmless $B$-meson decays are of crucial importance to deepen our
insights into the flavor structure of the standard model~(SM), the
origin of CP violation, the dynamics of hadronic decays, and to
search for any signals of new physics beyond the SM. The CP
violation, as an important part of $B$ physics, has been paid much
attention in recent years. Both mixing-induced and direct CP
violations have been observed in the neutral $B$ meson
decays~\cite{BB1,BB2,BB3,BB4,BB5}, which has provided a new window
for exploring the origin and mechanism of CP violation after the
establishment of indirect and direct CP violations in kaon
decays~\cite{K1,K2,K3,Wu01,Wu92}. The possible implications of
charmless B decays and their large CP violation have been
investigated in the recent papers~\cite{CLB,WZ1,WZ2}. In the SM, the
only source of CP violation in the SM is the single
Kobayashi-Maskawa phase~\cite{Kobayashi:1973fv} in the mixing matrix
that describes the charged current weak interaction of quarks.
However, physics beyond the SM are usually associated with new
sources of CP violation. For instance, rich sources of CP violation
can be induced from a single relative phase of vacuum in the simple
two Higgs doublet model with spontaneous CP violation(S2HDM or type
III 2HDM)~\cite{S2HDM}. The model can provide a natural explanation
for the CP violation in the SM and also lead to a new type of
CP-violating source with each quark and lepton carrying a nontrivial
CP-violating phase. Therefore, explorations of CP violation may well
indicate physics beyond the SM, or may be very helpful to
distinguish between various realizations of one particular kind of
new physics after the corresponding new physics particles have been
observed directly.

In addition to being served as a tool of looking for any new
physics, studying of CP violation can also be used to test various
factorization hypotheses, such as the ``naive" factorization
approach~(FA)~\cite{Wirbel}, the QCD factorization (QCDF)
\cite{M}, and the perturbation QCD method~(pQCD)~\cite{lihn}.
These methods have very different understandings of hadronic
$B$-meson decays: For the FA method, it cannot predict the direct
CP asymmetries properly due to the assumption of no strong
re-scattering in the final states; the pQCD generally predicts
large strong phases and big direct CP violations. Furthermore,
this approach also believes that annihilation diagrams are
important in non-leptonic two-body $B$-meson
decays~\cite{Keum:2000ph,Keum:2000ms}; while the QCDF favors small
direct CP violations in general because of the
$\alpha_s$-suppressed strong phases.

It is known that, in most cases of two-body $B$-meson decays, the
weak annihilation contribution carries different weak and strong
phases for the tree and penguin amplitudes, which is very important
for studying CP-violating observables. Meanwhile, the calculation of
annihilation contributions is interesting by itself, since it can
help us to understand the low energy dynamics of QCD involved in
heavy meson decays and the viability of the theoretical approaches
mentioned above. Motivated by these arguments, we shall investigate
in this paper the pure annihilation processes $\bar{B}^0\rightarrow
K^+K^-$, $K^{*\pm}K^\mp$, $K^{*+}K^{*-}$. These decay channels have
some interesting features: firstly, they are all pure annihilation
processes and studying these decay modes in the SM can serve as a
probe for the annihilation strength in hadronic $B$-meson decays;
secondly, the current experimental data on the decay $\bar B^0 \to
K^+K^-$~\cite{Barbar} has already provided a direct evidence for the
existence of the annihilation contributions. By comparing our
theoretical result for the branching ratio of the decay $\bar
{B}^0\rightarrow K^{+} K^{-}$ with the experimental data, we can
test the feasibility of the theoretical method and get a deeper
insight into the penguin annihilation and the $W$-exchanged
topologies in $B\rightarrow \pi\pi, \pi K$ decay
modes~\cite{BFRS,WZ2}; thirdly, their strong phases are calculable
and found to be big, which leads to large CP violation. The
resulting branching ratios for $\mathcal{B}(\bar {B}^0\rightarrow
K^{+} K^{-})$ is consistent with the current experimental
data~\cite{Barbar}; Finally, studying the polarization in the
$\bar{B}^0\rightarrow K^{*+} K^{*-}$ decay may also help us to
clarify whether annihilation contribution could resolve or alleviate
the polarization puzzles in $B\to \phi K^*$ decays as suggested in
Ref.~\cite{Kagan}.

In calculating the hard scattering kernel, we shall use the
Cornwall~\cite{Cornwall} prescription of gluon propagator by
introducing a dynamical mass of gluon to avoid enhancements in the
soft endpoint region. It is interesting to note that recent
theoretical~\cite{Alkofer} and phenomenological~\cite{S.Brodsky}
studies are now accumulating  supports for a softer infrared
behavior for the gluon propagator. Moreover, we will adopt the
Cutkosky rule~\cite{cutkosky} to deal with the physical-region
singularity caused by the on mass-shell quark propagators, which
then produce big imaginary parts from the kinematic region where
the virtual quark becomes on mass-shell. By applying  these two
``tricks", we observe that the main contributions to the decay
amplitudes come from the nonfactorizable diagrams, and the
CP-averaged branching ratios of $\bar B^0\rightarrow K^+K^-,
K^{+}K^{*-}, K^{-}K^{*+} $, and $K^{*+}K^{*-}$ are estimated by
using the QCD factorization to be about $2.02\times
10^{-8},4.23\times10^{-8}, 5.70\times10^{-8}$, and
$6.89\times10^{-8}$ for a given gluon mass $m_g=500$~MeV,
respectively. Moreover, big strong phases are predicted in these
decay modes, and hence the direct and mixing-induced CP violations
$C_{KK}$ and $S_{KK}$ are found to be very large in these decays
as the differences $\Delta C_{KK}$ and $\Delta S_{KK}$ are small
in $\bar{B}^0\rightarrow K^{\pm}K^{*\mp}$ decays. The predictions
may be tested in the more precise experiments at B-factory and the
LHC-b.

The paper is organized as follows. In section II, we first analysis
the relevant Feynman diagrams and then outline the necessary
ingredients for evaluating the CP asymmetries of the
$\bar{B}^0\rightarrow K^+K^-$, $K^{*\pm}K^\mp$, $K^{*+}K^{*-}$
decays. In section III, we present the approaches for dealing with
the physical-region singularities of gluon and quark propagators.
The numerical results of the CP-averaged branching ratios and large
strong phases are given in section IV. Finally, we discuss CP
asymmetries for those decay modes and give conclusions in sections V
and VI, respectively. The necessary input parameters and the
complete decay amplitudes for those decay modes are given in the
appendixes.

\section{Rephase-Invariant CP-Violating Observables}

Using the operator product expansion and renormalization group
equation, the low energy effective Hamiltonian for charmless
two-body $B$-meson decays can be written as~\cite{Buchalla}
\begin{eqnarray}
{\cal
H}_{eff}&=&\frac{G_{F}}{\sqrt{2}}\biggl\{V_{ub}V_{ud}^*[C_{1}(\mu)O_{1}(\mu)+
C_{2}(\mu)O_{2}(\mu)]
-V_{tb}V_{td}^*\sum\limits_{i=3}^{10}C_{i}(\mu)O_{i}(\mu)\biggl\}+h.c,
\end{eqnarray}
where $C_{i}(\mu)(i=1,\cdots,10)$ are the Wilson coefficient
functions which have been reliably evaluated to next-to-leading
logarithmic order. The effective operators $O_{i}$ are defined as
follows:
\begin{equation}
\begin{array}{ll}
O_{1}=(\bar{d}_{i}u_{i})_{V-A}(\bar{u}_{j}b_{j})_{V-A},
\hspace{1.5cm}
& O_{2}=(\bar{d}_{i}u_{j})_{V-A}(\bar{u}_{j}b_{i})_{V-A},\\
O_{3}=(\bar{d}_{i}b_{i})_{V-A}\sum\limits_{q}(\bar{q}_{j}
q_{j})_{V-A}, &
O_{4}=(\bar{d}_{i}b_{j})_{V-A}\sum\limits_{q}(\bar{q}_{j}
q_{i})_{V-A},\\
O_{5}=(\bar{d}_{i}b_{i})_{V-A}\sum\limits_{q}(\bar{q}_{j}
q_{j})_{V+A}, &
O_{6}=(\bar{d}_{i}b_{j})_{V-A}\sum\limits_{q}(\bar{q}_{j}
q_{i})_{V+A},\\
O_{7}=\frac{3}{2}(\bar{d}_{i}b_{i})_{V-A}\sum\limits_{q}e_{q}
(\bar{q}_{j}q_{j})_{V+A}, &
O_{8}=\frac{3}{2}(\bar{d}_{i}b_{j})_{V-A}\sum\limits_{q}e_{q}
(\bar{q}_{j}q_{i})_{V+A},\\
O_{9}=\frac{3}{2}(\bar{d}_{i}b_{i})_{V-A}\sum\limits_{q}e_{q}
(\bar{q}_{j}q_{j})_{V-A}, &
O_{10}=\frac{3}{2}(\bar{d}_{i}b_{j})_{V-A}\sum\limits_{q}e_{q}
(\bar{q}_{j}q_{i})_{V-A}.
\end{array}
\end{equation}
Here $i$ and $j$ are SU(3) color indices, $q$ denotes all the
active quarks at the scale $\mu={\cal O}(m_{b})$, i.e.,
$q=u,d,s,c,b.$

With the effective Hamiltonian, calculations of the leading order
amplitudes for $\bar{B}^0\rightarrow K^+K^-, K^{*\pm}K^\mp,
K^{*+}K^{*-}$ decays are straightforward. However, due to the
conservation of the vector current and partial conservation of the
axial-vector current, the leading order amplitudes will vanish in
the limit $m_u,~m_s \to 0$. In order to probe the annihilation
strength and discuss CP violation in these processes, we have to
consider the next-to-leading order~($\alpha_s$ order)
contributions.

\begin{figure}[t]
\begin{center}
\scalebox{0.7}{\epsfig{file=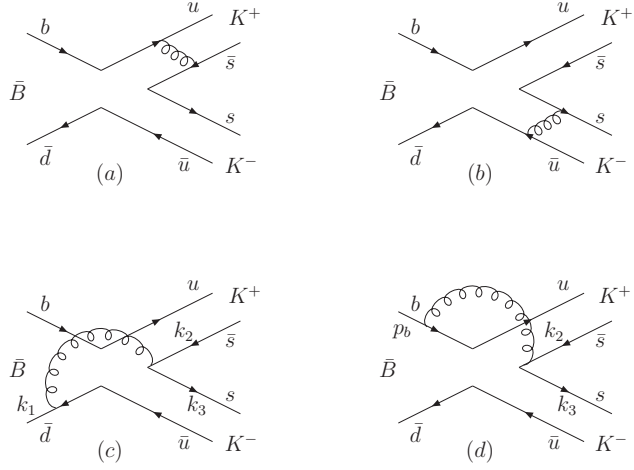}} \caption{\small The
$W$-exchanged annihilation diagrams for $\bar{B}^0\rightarrow
K^{+}K^{-}$ decay.} \label{tree}
\end{center}
\end{figure}

\begin{figure}[t]
\begin{center}
\scalebox{0.7}{\epsfig{file=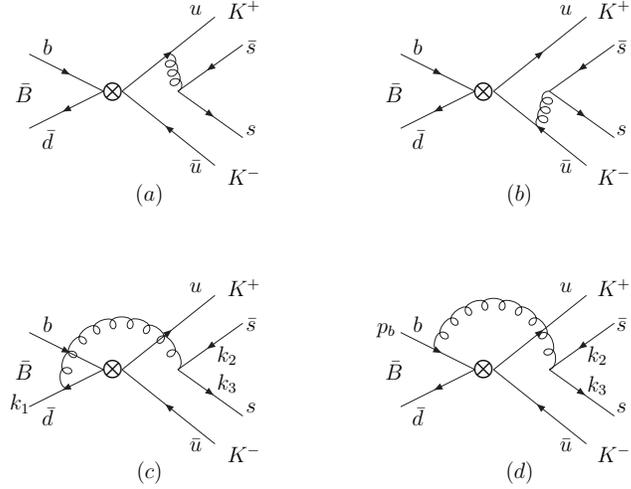}} \caption{\small The
penguin annihilation diagrams for $\bar B^0\rightarrow K^{+}K^{-}$
decay.}\label{penguin}
\end{center}
\end{figure}

Up to the $\alpha_s$ order, the relevant Feynman diagrams
contributing to the $\bar B^0\rightarrow K^+K^-$ decay~(the
corresponding diagrams for $K^{*\pm}K^\mp$ and $K^{*+}K^{*-}$
modes are the same) are depicted in Fig.~\ref{tree}~
(corresponding to the $W$-exchanged annihilation diagrams) and
Fig.~\ref{penguin}~(corresponding to the penguin annihilation
diagrams and including the case with the exchange of
$u\leftrightarrow s$). For the factorizable diagrams (a) and (b)
in Figs.~\ref{tree} and~\ref{penguin},  their contributions cancel
each other both in the $W$-exchanged and in the penguin
annihilation diagrams, so that the non-factorizable contributions
will dominate the decay, which can be obtained by calculating the
amplitudes of diagrams (c) and (d) in Figs.~\ref{tree}
and~\ref{penguin}. Moreover, it is noted that these properties
hold in all of these decay modes.


 The direct CP violation occurs only if there are two contributing
amplitudes with non-zero relative weak and strong phases. The weak
phase difference can arise from the interference of amplitudes
from various tree(current-current) and penguin diagrams. From the
Feynman diagrams in Figs.~\ref{tree} and~\ref{penguin}, we can see
that in these decays there are two kinds of CKM elements,
$V_{ub}V_{ud}^*$ from tree operators and $V_{tb}V_{td}^*$ from
penguin ones, which will induce weak phase difference and hence CP
violation. The total decay amplitudes for $B^0(\bar B^0)\to
K^+K^-$ mode can be written as
\begin{eqnarray}
{\cal A}({B}^0\rightarrow K^{+}K^{-})=V_{ub}^*V_{ud}{\cal
A}_T-V_{tb}^*V_{td}{\cal A}_P=V_{ub}^*V_{ud}{\cal
A}_T[1+ze^{i(\alpha+\delta)}],\label{b0} \\
{\cal A}(\bar{B}^0\rightarrow K^{+}K^{-})=V_{ub}V_{ud}^*{\cal
A}_T-V_{tb}V_{td}^*{\cal A}_P=V_{ub}V_{ud}^*{\cal
A}_T[1+ze^{i(-\alpha+\delta)}]\label{bbar0},
\end{eqnarray}
where $\alpha=\arg[-V_{tb}^*V_{td}/V_{ub}^*V_{ud}]$,
$z=|V_{tb}^*V_{td}/V_{ub}^*V_{ud}||{\cal A}_P/{\cal A}_T|$, which
indicates the interference strength between the annihilation
amplitudes from penguin and tree operators, and $\delta=\arg({\cal
A}_P/{\cal A}_T)$ is the relative strong phase between the penguin
and the tree annihilation amplitudes. A similar consideration can be
applied to the $ B^0\rightarrow K^{\pm}K^{*\mp}$ and $B^0\rightarrow
K^{*+}K^{*-}$ decays. The resulting decay amplitudes by using QCD
factorization are given in Appendix B.

For neutral $B$-meson decays, the time-dependent CP asymmetries
are defined as
\begin{equation}
{\cal  A}_{CP}(t) = \frac{\Gamma(B^0(t) \to f )
 -\Gamma(\bar{B}^0(t)
\to \bar{f} )}{ \Gamma({B}^0(t) \to {f}) + \Gamma(\bar B^0(t) \to
\bar f)}.\label{eq:acp0}
\end{equation}

When the final state is a CP eigenstate, such as the $B^0(\bar
{B}^0) \to K^+K^-$ decay, the time-dependent CP asymmetries can be
written as
\begin{equation}
{\cal  A}_{CP}(t)=C_{KK}\cos(\triangle mt)+ S_{KK}\sin(\triangle
mt), \label{acp1}
\end{equation}
where $\Delta m$ is the mass difference of the two eigenstates of
$B_d$ mesons. $C_{KK}$ and $S_{KK}$ are parameters describing the
direct and the mixing-induced CP violation, respectively. Both of
them depend on the CKM and hadronic matrix elements
\begin{equation}
C_{KK}=\frac{1-|\lambda_{CP}|^2}{1+|\lambda_{CP}|^2},\hspace{1cm}
S_{KK}=\frac{-2Im(\lambda_{CP})}{1+|\lambda_{CP}|^2},
\end{equation}
with
\begin{equation}
\lambda_{CP}=\frac{V^*_{tb}V_{td}\langle K^+K^-|H_{eff}|\bar
B^0\rangle}{V_{tb}V^*_{td}\langle K^+K^-|H_{eff}|B^0\rangle}.
\end{equation}
From Eqs.~(\ref{b0}) and~(\ref{bbar0}), the CP-violating
parameters $C_{KK}$ and $S_{KK}$ can be expressed explicitly as
\begin{equation}
C_{KK}=\frac{-2z\sin\alpha\sin\delta}{1+2z\cos\alpha\cos\delta+z^2},
\hspace{1cm}
S_{KK}=\frac{-\sin2\alpha-2z\sin\alpha\cos\delta}{1+2z\cos\alpha\cos\delta+z^2},
\label{acp2}
\end{equation}
which shows that both the direct and the mixing-induced CP violation
depend not only on the strong phase $\delta$ but also on the
magnitudes of $z$. Thus, when the contributions of different weak
decay amplitudes are comparable each other, there will be a high
likelihood for observable CP-violating asymmetries.

For the case of $B\to K^{*\pm}K^\mp$ decays, as the final state is
not an CP eigenstate, the CP-violating asymmetries become
complicated. There are in general four decay amplitudes which can
be expressed as
\begin{equation}
\begin{array}{ll}
g=\langle K^+K^{*-}|H_{eff}|B\rangle, \hspace{2cm} h=\langle
K^+K^{*-}|H_{eff}|\bar B\rangle,\\ \bar g=\langle
K^-K^{*+}|H_{eff}|\bar B\rangle, \hspace{2cm} \bar h=\langle
K^-K^{*+}|H_{eff}|B\rangle.\label{eq:acp2}
\end{array}
\end{equation}
Following the discussions in Ref.~\cite{PW}, there exist in
general four rephase-invariant parameters $a_{\epsilon'}$,
$a_{\bar\epsilon'}$, $a_{\epsilon+\epsilon'}$,
$a_{\epsilon+\bar\epsilon'}$ which characterize the CP
asymmetries. We may redefine the following four parameters
\begin{equation}
\begin{array}{ll}
C_{KK}=\frac{1}{2}(a_{\epsilon'}+a_{\bar\epsilon'}), \qquad\,
&\Delta C_{KK}=\frac{1}{2}(a_{\epsilon'}- a_{\bar\epsilon'}),\\
S_{KK}=\frac{1}{2}(a_{\epsilon+\epsilon'}+a_{\epsilon+\bar\epsilon'}),
\qquad\,&\Delta
S_{KK}=\frac{1}{2}(a_{\epsilon+\epsilon'}-a_{\epsilon+\bar\epsilon'}),
\end{array}
\end{equation}
with
\begin{eqnarray}
&& a_{\epsilon'}=\frac{|g|^2-|h|^2}{|g|^2+|h|^2}, \qquad\,
a_{\bar\epsilon'}=\frac{|\bar h|^2-|\bar g|^2}{|\bar h|^2+|\bar
g|^2},\nonumber\\
&& a_{\epsilon +\epsilon'}=\frac{-2\,{\rm Im}
(h/g)}{1+|h/g|^2},\qquad\, a_{\epsilon+\bar \epsilon'}=\frac{-2\,
{\rm Im} (\bar g/ \bar h)}{1+|\bar g/\bar h|^2}.
\end{eqnarray}
For final state being CP eigenstate, one has $\Delta C_{KK} =0$
and $\Delta S_{KK} =0$.

As for the $B\to K^{*+}K^{*-}$ decay mode, since the total
amplitudes are dominated by the longitudinal ones, which can be
seen below, one can evaluate its CP asymmetries in the same way as
for the $B\to K^+K^-$ decay.

\section{Treatments for Physical-region Singularities of Gluon and Quark Propagators  }

To perform a numerical calculation, the QCD factorization approach
may be used to evaluate the amplitudes of $\bar{B}^0\rightarrow
K^+K^-, K^{*\pm}K^\mp, K^{*+}K^{*-}$ decays. The details are
presented in Appendix B. In Eqs.~(\ref{pp1})--(\ref{vv}), one will
encounter the endpoint divergence, which is the most difficult part
to deal with in the annihilation diagrams within the QCD
factorization framework. Instead of the widely used treatment
$\int^1_0 \frac{dy}{y}\to X_{A}=(1+\varrho_A
e^{i\varphi})\ln\frac{m_B}{\Lambda_h} $ in the
literature~\cite{bn,M.Beneke,DU}, we shall use an effective gluon
propagator~\cite{Cornwall,Yang} to treat the infrared divergence in
the soft endpoint region
\begin{equation}
\frac{1}{k^{2}}~\Rightarrow~\frac{1}{k^{2}+M_g^2(k^{2})},
\hspace{1cm}M_g^2(k^{2})=m_g^2 \biggl
[\frac{\ln(\frac{k^{2}+4m_g^2}{\Lambda^{2}})}
{\ln(\frac{4m_g^2}{\Lambda^{2}})}\biggl ]^{-\frac{12}{11}}.
\end{equation}
The typical values $m_{g}=(500\pm200$)~MeV, and
$\Lambda=\Lambda_{QCD}$=250~MeV will be taken in our numerical
calculations. Use of this gluon propagator is supported by the
lattice~\cite{Williams} and the field theoretical
studies~\cite{Alkofer}, which have shown that the gluon propagator
is not divergent as fast as $\frac{1}{k^{2}}$.

After giving the treatments for the infrared divergence arising
from the gluon propagator, we now turn to show how to deal with  a
physical-region singularity of the on mass-shell quark
propagators. It can be easily checked that each Feynman diagram
contributing to a given matrix element is entirely real unless
some denominators vanish with a physical-region singularity, so
that the $i\epsilon$ prescription for treating the poles becomes
relevant. In other words, a Feynman diagram will yield an
imaginary part for decay amplitudes when the virtual particles in
the diagram become on mass-shell, thus the diagram may be
considered as a genuine physical process. The Cutkosky
rules~\cite{cutkosky} give a compact expression for the
discontinuity across the cut arising from a physical-region
singularity. When applying the Cutkosky rules to deal with a
physical-region singularity of quark propagators, the following
formula holds
\begin{eqnarray}
\frac{1}{(k_1-k_2-k_3)^2+i\epsilon}&=&P\biggl[\frac{1}{(k_1-k_2-k_3)^2}
\biggl]-i\pi\delta[(k_1-k_2-k_3)^2],\label{quarkd}\\
\frac{1}{(p_b-k_2-k_3)^2-m_b^2+i\epsilon}&=&P\biggl[\frac{1}
{(p_b-k_2-k_3)^2-m_b^2}\biggl]-i\pi\delta[(p_b-k_2-k_3)^2-m_b^2],
\label{quarkb}
\end{eqnarray}
where $P$ denotes the principle-value prescription. The role of the
$\delta$ function is to put the particles corresponding to the
intermediate state on their positive energy mass shell, so that in
the physical region, the individual diagrams satisfy the unitarity
condition.  Eqs.~(\ref{quarkd}) and~(\ref{quarkb}) will be applied
to the quark propagators $D_d$ and $D_b$ in
Eqs.~(\ref{pp1})--(\ref{vv}), respectively. It is then seen that the
big imaginary parts arise from the virtual quarks~(d, b) across
their mass shells as physical-region singularities. In fact, the
above imaginary parts are among the main sources of strong phases
for the $\bar{B}^0\rightarrow K^+K^-, K^{*\pm}K^\mp, K^{*+}K^{*-}$
decays as discussed in the perturbative QCD
approach~\cite{Keum:2000ms,Lu:2000em}.

\section{Decay Amplitudes and Large Strong Phases }

Using the relevant input parameters listed in Appendix A, we can
calculate the tree and the penguin annihilation amplitudes for each
decay mode and their corresponding numerical results for the
quantities $z$ and $\delta$, which are presented in
Table~\ref{strong phase}. With these considerations, the CP-averaged
branching ratios for these decay modes are given in Table~\ref{br1}.
In this table, we present our ``default results" along with detailed
error estimates corresponding to the different theoretical
uncertainties listed in Appendix A. The first error refers to the
variation of the dynamical gluon mass, while the second one the
uncertainty due to the CKM parameters $A, \lambda, \bar{\rho}$, and
$\bar{\eta}$. Finally, the last error originate from the uncertainty
due to the meson decay constants and the parameter $\mu_K$.

\begin{table}
\caption{ \label{strong phase} The tree annihilation amplitudes
${\cal A}_T$, the penguin annihilation amplitudes ${\cal A}_P$,
magnitudes of $z$ and the strong phase $\delta$ for
$B^0(\bar{B}^0)\rightarrow K^+K^-, K^{*\pm}K^\mp, K^{*+}K^{*-}$
decays. Here we give only the case when $m_g=500$ MeV.}
\begin{center}
\begin{tabular}{lccccc}\hline \hline
Decay mode&${\cal A}_T$&${\cal A}_P$&$z$&$\delta$\\
\hline\hline
 $B^{0}(\bar B^{0}) \to K^{+} {K}^- $
 &~~ $0.0273-0.0321i$&~~$-0.0114+0.0045i$&~~$0.65$&~~$-153^\circ$ \\
 \hline$B^{0}(\bar B^{0}) \to K^{-}K^{*+}$
 & ~~$0.0467-0.0371i$&~~$0.0187-0.0005i$&~~$0.70$&~~$37^\circ$\\
 $B^{0}(\bar B^{0}) \to K^+K^{*-}$
 &~~ $0.0412-0.0464i$&~~$0.0333-0.0092i$&~~$1.14$&~~$33^\circ$\\\hline
 $ B^0(\bar B^{0})\to K^{*+} K^{*-} $
 &~~ $0.0634-0.0543i$&~~$-0.0168+0.0047i$&~~$0.47$&~~$-155^\circ$\\
\hline\hline
\end{tabular}
\end{center}
\end{table}
\begin{table}
\caption{ \label{br1} The CP-averaged branching ratios~(in units of
$10^{-8}$) of $B^0(\bar{B}^0) \rightarrow K^+K^-$, $K^{*\pm}K^\mp$,
$K^{*+}K^{*-}$ decays. The theoretical errors shown from left to
right correspond to the uncertainties referred to as ``gluon mass",
``CKM parameters", and ``decay constants and the parameter $\mu_K$"
as specified in the text.}
\begin{center}
\begin{tabular}{lcccc}\hline \hline
Decay mode&~~$Br$&~~$\Gamma_T/\Gamma$&Exp.\\
\hline\hline
 $B^{0}(\bar B^{0})\to K^{+} {K}^- $
 &~~ $2.02^{\,+3.21\,+0.83\,+1.01}_{\,-0.81\,-0.54\,-0.44}$
 & ~~$-$&~~
 $(4\pm15\pm8)\times10^{-8}$~\cite{Barbar} \\
 \hline$B^{0}(\bar B^{0}) \to K^{-}K^{*+}$
  &~~$5.70^{\,+2.08\,+2.26\,+\,1.11}_{\,-3.01\,-1.71\,-0.92}$ &~~$-$&~~$-$\\
 $B^{0}(\bar B^{0}) \to K^+K^{*-}$
  &~~$4.23^{\,+2.29\,+1.63\,+0.94}_{\,-2.05\,-1.21\,-0.81}$ &~~$-$&~~$-$\\ \hline
 $ B^0(\bar B^{0})\to K^{*+} K^{*-} $
  &~~$6.89^{\,+1.78\,+2.87\,+2.04}_{-1.50\,-2.19\,-1.58}$
  &~~$0.01$&~~
 $<1.41\times10^{-4}$~\cite{s.eidelman}\\
  \hline\hline\end{tabular}
\end{center}
\end{table}

From the numerical results given in Tables.~\ref{strong phase} and
\ref{br1}, we arrive at the following observations:
\begin{itemize}
\item The strong phase associated with the tree amplitude ${\cal
A}_T$ is about $\delta \simeq -45^\circ$, while the imaginary part
of the penguin amplitude ${\cal A}_P$ is comparatively small;
moreover, from the numerical results of the magnitude $z$, we can
see that the penguin annihilation amplitudes are comparable to the
tree ones, so that a large interference effect between the tree and
the penguin annihilation amplitudes occurs. Combining these two
ingredients, it is expected that there exist large CP violations in
these decay modes, which will be shown below.

\item For a given dynamical gluon mass, the CP-averaged branching
ratios of these decay modes follow the pattern:
\begin{equation}
\mathcal{B}(\bar B^0\rightarrow K^{*+}K^{*-})> \mathcal{B}(\bar
B^0\rightarrow K^{*\pm}K^\mp)> \mathcal{B}(\bar B^0\rightarrow
K^+K^-),
\end{equation}
which is due to the larger vector-meson decay constant
$f_{K^*}>f_K$ for one vector meson in the final state or the
larger spin phase space available for two final-state vector
mesons.

\item For the $\bar B^0\to K^+K^-$ decay, the present obtained
result is consistent with the experimental data~\cite{Barbar}, and
also in agreement with the one given in~\cite{bn}:
$\mathcal{B}(\bar{B}^0\to
K^{+}K^{-})=(0.013^{+0.005+0.008+0.087}_{-0.005-0.005-0.011})\times
10^{-6}$, when considering the huge uncertainties caused in the
treatment of infrared divergence $\int^1_0 \frac{dy}{y}\to
X_{A}=(1+\varrho_A e^{i\varphi})\ln\frac{m_B}{\Lambda_h} $. The
present theoretical errors mainly originate from the variation of
dynamical gluon mass, while our theoretical result alleviates the
dependence of input parameters when compared to the one given by
Ref.~\cite{bn}.

~~~Another significant error comes from the parameter $\mu_K$, as
the decay amplitudes include $\mu_K^2$ factor when considering the
twist-3 wave functions contributions. Moreover, the present
central value is smaller than the one given by the pQCD~\cite{kk}:
$Br(B^0\to K^+K^-)=4.6\times 10^{-8}$, which has used a bigger
parameter $\mu_K$. If we also choose $\mu_K=2.22$~GeV just as the
pQCD does, and a dynamical gluon mass with $m_g=420$~MeV, we can
get the same result as the pQCD prediction. Namely, if choosing
smaller $\mu_K$, to get the same result as the pQCD prediction,
one has to choose a smaller $m_g$. In general, with the parameter
$\mu_K$ fixed at the value 1.4, 1.8 and 2.2~GeV, we can get the
same result as the pQCD by taking $m_g$ to be $m_g=340$, 370, and
420~MeV, respectively. Anyway, our prediction is consistent with
the pQCD result after taking into account the theoretical
uncertainties.

\item For the $\bar B^0\to K^{\pm}K^{*\mp}$ decays, from the numerical
results we can see that the main theoretical errors originate from
dynamical gluon mass and CKM parameters. In addition, we predict
that $\mathcal{B}(\bar B^0\rightarrow K^+K^{*-})\neq
\mathcal{B}(\bar B^0\rightarrow K^-K^{*+})$. The central values are
larger than the ones given in Ref.~\cite{bn}:
$\mathcal{B}(\bar{B}^0\to
K^{\pm}K^{*\mp})=(0.014^{+0.007+0.010+0.106}_{-0.006-0.006-0.012})\times
10^{-6}$, but they remain consistent with each other within the
uncertainties.

\item For the $\bar B^0\to K^{*+}K^{*-}$ decay, only an upper
limit at $90\%$ confidence level has been
reported~\cite{s.eidelman}
\begin{equation}
\mathcal{B}(\bar{B}^0\rightarrow K^{*+}K^{*-})<1.41\times10^{-4}.
\end{equation}
Obviously, the present numerical result is far below the
experimental data. It is noted that the branching ratio of this
decay channel is less sensitive to the dynamics gluon mass, and the
theoretical errors mainly come from CKM parameters. It is also noted
that $99\%$ of the branching ratio comes from the longitudinal part,
and the result is consistent with that given by the pQCD
method~\cite{jinzhu}.
\end{itemize}

\section{Large CP Violation}

We now turn to discuss the CP asymmetries in the
$B^0(\bar{B}^0)\rightarrow K^+K^-, K^{*\pm}K^\mp, K^{*+}K^{*-}$
decays. As there are the big strong phases and large interference
effects in these decay modes, large CP-violating asymmetries are
expected. It is very reasonable to neglect the transverse
contribution and consider only the longitudinal part in $\bar
B^0\to K^{*+}K^{*-}$ decay, since the transverse polarization
fraction provides only $1\%$ contribution to the total branching
ratio of this mode. Thus we can discuss the CP asymmetries in the
$K^+K^-$ and $K^{*+}K^{*-}$ decay modes by the same manner.

\begin{figure}[htbp]
\begin{center}
\begin{tabular}{cc}
\scalebox{0.7}{\epsfig{file=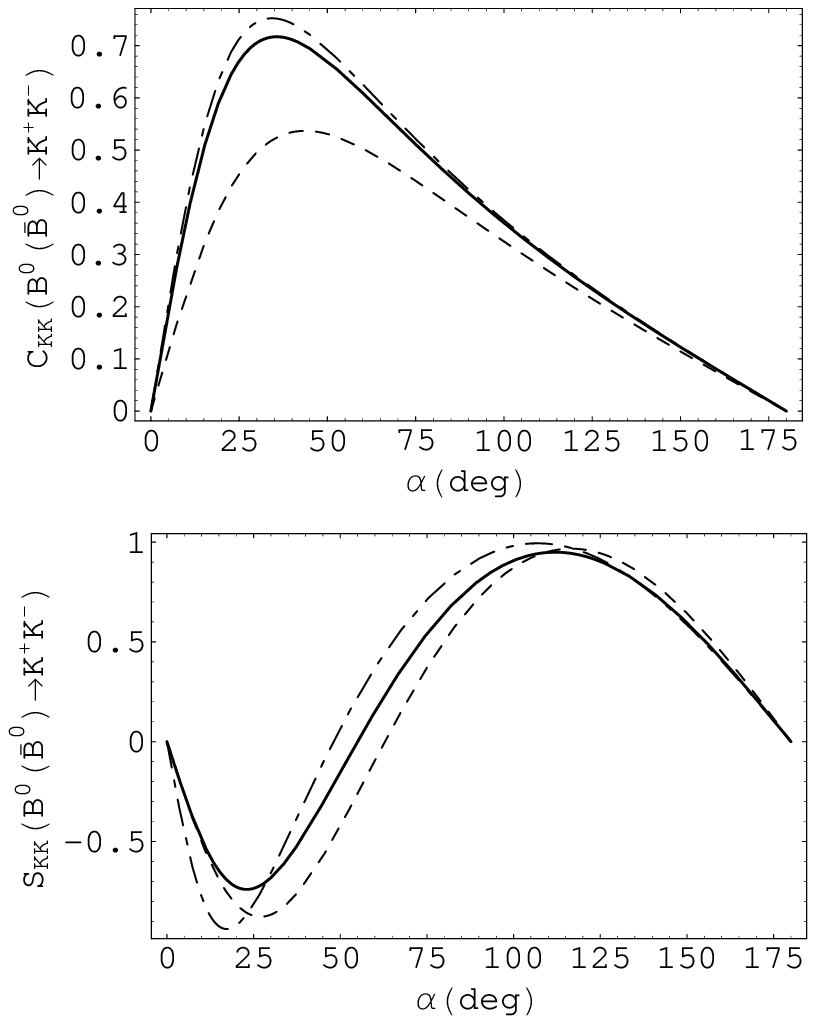}} & \scalebox{0.7}
{\epsfig{file=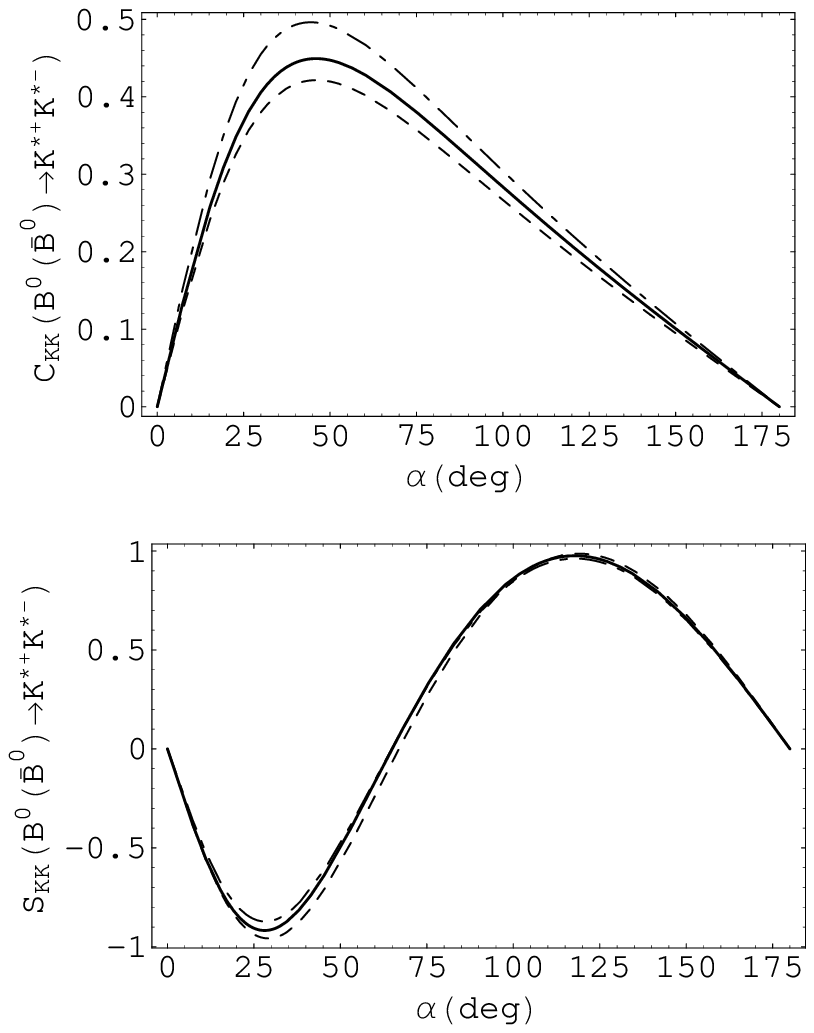}}
\end{tabular}
\caption{\small The CP violation parameters $C_{KK}$ and $S_{KK}$
for $B^0(\bar{B}^0)\rightarrow K^+K^-, K^{*+}K^{*-}$ decays as
functions of the weak phase $\alpha$~(in degree). The dash-dotted,
solid, and dashed lines correspond to $m_g=300$~MeV, $500$~MeV,
and $700$~MeV, respectively.} \label{ppvv}
\end{center}
\end{figure}

\begin{figure}[htbp]
\begin{center}
\begin{tabular}{cc}
\scalebox{0.7}{\epsfig{file=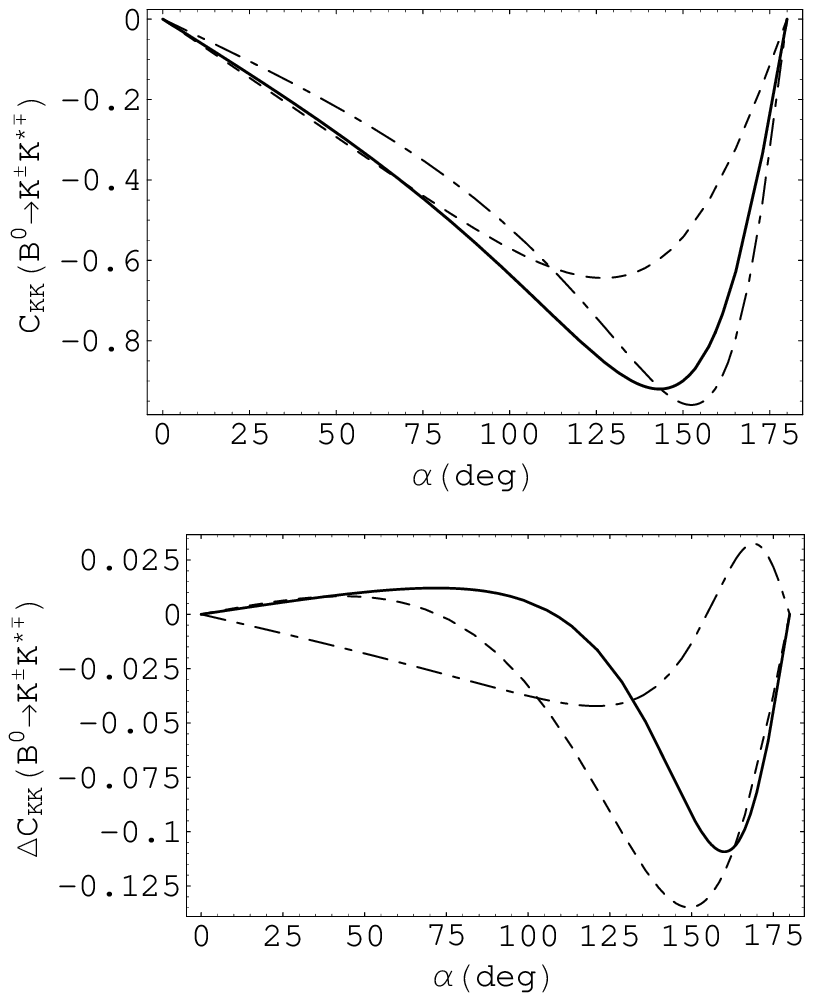}} & \scalebox{0.7}
{\epsfig{file=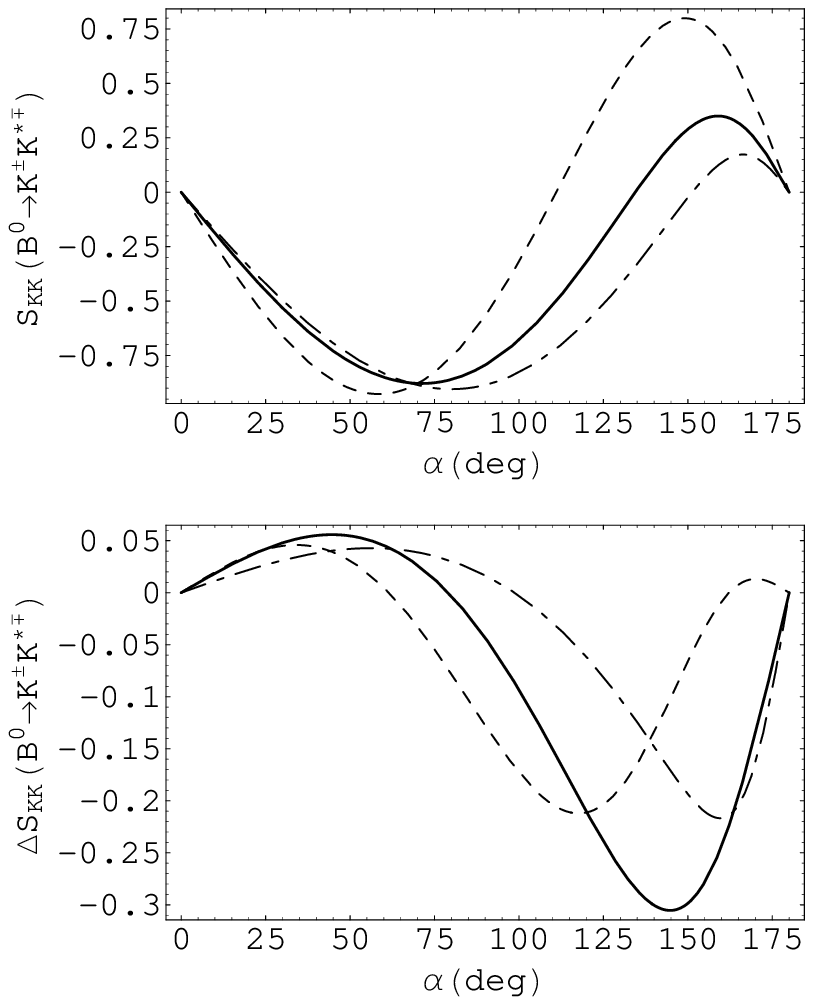}}
\end{tabular}
\caption{\small The same as Fig.~\ref{ppvv} but for
$B^0(\bar{B}^0) \rightarrow K^{\pm}K^{*\mp}$ decays.}\label{pv}
\end{center}
\end{figure}

Using the relevant formulas provided in the previous section, we are
able to calculate the CP violations in the $\bar{B}^0\rightarrow
K^+K^-, K^{*\pm}K^\mp, K^{*+}K^{*-}$ decays. Firstly, in
Table~\ref{cp}, we present our predictions for the rephase-invariant
CP-violating observables and their theoretical errors in these
decays. Secondly, taking the CKM angle $\alpha$ as a free parameter,
the dependence of CP-violating parameters on the angle $\alpha$ is
shown in Figs.~\ref{ppvv} and~\ref{pv}.

\begin{table}[htbp]
\caption{\label{cp} The CP asymmetries for
$B^0(\bar{B}^0)\rightarrow K^+K^-, K^{*\pm}K^\mp, K^{*+}K^{*-}$
decays with the same error resources as in Table~\ref{br1}.}
\begin{center}
\doublerulesep 0.8pt \tabcolsep 0.1in
\begin{tabular}{lccccc}\hline \hline
  mode& $C_{KK}$ & $\Delta C_{KK}$& $S_{KK}$& $\Delta S_{KK}$&
\\\hline\hline
 $K^{+} {K}^- $
 & $0.39^{\,+0.00\,+0.04\,+0.01}_{\,-0.04\,-0.04\,-0.01}$&$0.00$
 &$0.86^{\,+0.09+\,0.04\,+0.02}_{\,-0.05\,-0.08\,-0.05}$&$0.00$\\
 $K^{\pm}K^{*\mp}$
 &$-0.59^{\,+0.11\,+0.05\,+0.01}_{\,-0.00\,-0.07\,-0.01}$
 &$0.01^{\,+0.00\,+0.01\,+0.01}_{\,-0.04\,-0.01\,-0.01}$&
 $-0.74^{\,+0.29\,+0.02\,+0.06}_{\,-0.12\,-0.08\,-0.04}$&
 $-0.07^{\,+0.08\,+0.06\,+0.06}_{\,-0.08\,-0.06\,-0.06}$\\
 $K^{*+} K^{*-} $
 &$0.30^{\,+0.02\,+0.05\,+0.00}_{\,-0.01\,-0.05\,-0.00}$&$0.00$
 &$0.78^{\,+0.01\,+0.09\,+0.00}_{\,-0.01\,-0.10\,-0.00}$&$0.00$
 \\\hline\hline
\end{tabular}
\end{center}
\end{table}



From Table~\ref{cp} and Figs.3-4, we come to the following
observations:
\begin{itemize}

\item[]{(i)} For the $\bar B^0\to K^+K^-$ decay, due to the
large strong phase $\delta$ and large magnitude $z\simeq 0.65$, the
direct and mixing-induced CP-violating parameters $C_{KK}$ and
$S_{KK}$ are found to be quite large. The direct CP asymmetry is
consistent with the one given by pQCD~\cite{kk}. It is also seen
that the CP asymmetry parameters $C_{KK}$ and $S_{KK}$ are not
sensitive to the choice of the dynamical gluon mass, and the main
theoretical errors originate from CKM parameters.

\item[]{(ii)} For the $\bar B^0\to K^{*+}K^{*-}$ decay, as it is
dominated by the longitudinal part, the CP asymmetries in this
decay mode have the same manner as the one in the $\bar B^0\to
K^+K^-$ decay. Thus we arrive at the similar conclusions as the
ones for the $\bar B^0\to K^+K^-$ decay.

\item[]{(iii)} In contrast to the $\bar B^0\to K^+K^-$,
$K^{*+}K^{*-}$ decays, the strong phases $\delta$ in the $\bar
B^0\to K^{\pm}K^{*\mp}$ decay modes have opposite signs, so that
the sign of the CP asymmetry parameter $C_{KK}$  is also opposite
to the ones in the $K^+K^-, K^{*+}K^{*- }$ decay modes. In
addition, the rephase-invariant parameters $C_{KK}$ and $S_{KK}$
in the $\bar B^0\to K^{\pm}K^{*\mp}$ decay modes also characterize
large CP violation as the parameters $\Delta C_{KK}$ and $\Delta
S_{KK}$ are small. Note that only the mixing-induced CP violation
$S_{KK}$ is sensitive to the dynamical gluon mass.

\item[]{(iv)} It is seen from figures~\ref{ppvv} and \ref{pv}
that all CP-violating parameters have a strong
dependence on the weak angle $\alpha$. So these observables may be
used to determine the range of the angle $\alpha$ in future more
precise experiments.
\end{itemize}

\section{Conclusions}

In summary, we have calculated the strong phases, the CP-averaged
branching ratios, and the CP asymmetries for the pure annihilation
decays $\bar B^0\rightarrow K^{+}K^{-}, K^{\pm}K^{*\mp}$, and
$K^{*+}K^{*-}$ within the standard model. It has been shown that
the nonfactorizable contributions dominate these decays, and the
contributions of the penguin diagrams are comparable to that of
the $W$-exchanged diagrams; the CP-averaged branching ratios of
these decay modes are at the order of $10^{-8}\sim 10^{-7}$, and
these small branching ratios predicted in the SM make them
sensitive to any new physics beyond the SM. Of particular, as
there are big strong phases and large interference effects between
the tree and the penguin annihilation amplitudes, the CP-violating
parameters $C_{KK}$ and $S_{KK}$ have been predicted to be large
in these decay modes.  It has been seen that the CP-violating
parameters have a strong dependence on the weak phase $\alpha$,
but they are not sensitive to the dynamical gluon mass except the
mixing-induced CP violation in the $\bar B^0\rightarrow
K^{*+}K^{-}, K^{+}K^{*-}$ decays. The resulting branching ratio
$Br(\bar B^0 \to K^+K^-)$ is consistent with the current
experimental data. It is then expected that the predicted CP
asymmetries should be reasonable.

In this paper, we have adopted the Cornwall prescription for the
gluon propagator with a dynamical mass to avoid the endpoint
infrared divergence. Note that when the intrinsic mass is
appropriately introduced, it may not spoil the gauge symmetry as
shown recently in the symmetry-preserving loop
regularization~\cite{LR}. Meanwhile, we have also applied the
Cutkosky rules to deal with the physical-region singularity of the
on-mass-shell quark propagators. As a consequence, it produces the
big imaginary parts which are very important for understanding large
CP violations. The combination of the two treatments for the
endpoint infrared divergence of gluon propagator and the
physical-region singularity of the quark propagators enables us to
obtain, by using the QCD factorization approach~\cite{M}, reasonable
results which are consistent with the ones~\cite{kk,jinzhu} obtained
by using the perturbative QCD approach\cite{lihn}. However,
different from the treatment of perturbative QCD approach, where the
Sudakov factors have been used to avoid the endpoint divergence. As
a consequence, it was shown that the pQCD predictions are
insensitive to the choice of Sudakov factors and to the dependence
of the impact parameter b~\cite{Keum:2000ph}.

It is noted that our present predictions for branching ratios
depend on the dynamics gluon mass which plays the role of IR
cut-off, such a dependence should in general be matched from the
nonperturbative effects in the transition wave functions. However
this is not the case in general for direct CP violation. With the
experimental and theoretical improvements, this quantity could be
fitted from a well measured pure annihilation decay mode, and then
expanded to other decays. In order to check the validity of the
gluon-mass regulator method adopted to deal with the endpoint
divergence, we plan to extend this method to the vertex
corrections and hard spectator interactions for other charmless
$B$-meson decays. We expect that these corrections are
independence of the dynamical gluon mass, which is being under
investigation. Anyhow, the treatment presented in this paper may
enhance its predictive power for analyzing the charmless
non-leptonic $B$-meson decays.

\acknowledgments This work was supported in part by the National
Science Foundation of China (NSFC) under the grant 10475105,
10491306, and the Project of Knowledge Innovation Program (PKIP) of
Chinese Academy of Sciences.

\section*{Appendix A: Input parameters}

The theoretical predictions in our calculations depend on many
input parameters, such as the Wilson coefficient functions, the
CKM matrix elements, the hadronic parameters, and so on. We
present all the relevant input parameters as follows.

The next-to-leading order Wilson coefficient functions~(at
$\mu=m_b/2$) in the NDR scheme for $\gamma_5$~\cite{DU} have been
used with the following numerical values
\begin{equation}
\begin{array}{llll}
  C_1 =1.130, & C_2 =-0.274, & C_3 =0.021,& C_4= -0.048,\\
  C_5=0.010,& C_6= -0.061,& C_7= -0.005/128,& C_8=0.086/128,\\
  C_9=-1.419/128,& C_{10}=0.383/128.
\end{array}\label{ci}
\end{equation}

For the $B$ meson wave function, we have taken the following
results~\cite{h.y.cheng}
\begin{equation}
\Phi^B_1(\bar\rho)=N_B\bar\rho^2(1-\bar\rho)^2
\textrm{exp}\left[-\frac{1}{2} \left(\frac{\bar\rho
m_B}{\omega_B}\right)^2\right],
\end{equation}
with $\omega_B=0.25$~GeV, and $N_B$ being a normalization constant.
For the light cone wave functions of light mesons, we use the
asymptotic form as given in Refs.~\cite{M.T,Genon,p.ball}:
\begin{eqnarray}
\Phi_K(u)&=&\Phi_{\parallel}(u)=\Phi_{\perp}(u)=6u\bar u, \qquad
{\rm twist-2~~LCDAs},\nonumber\\
\phi_\sigma(u)&=&g^{(a)}_\perp(u)=h^{(s)}_\parallel(u)=6u\bar u,\nonumber\\
g_\perp^{(v)}(u) &=& \frac{3}{4}\left[1+(u-\bar u)^2\right],\nonumber\\
h_\parallel^{(t)}(u) &=& 3-12u\bar u,~~ \phi_k(u)=1, \qquad {\rm
twist-3~~LCDAs},
\end{eqnarray}
with $\bar u=1-u$.

For the other parameters used in our calculations, we list as
follows~\cite{s.eidelman}
\begin{equation}
\begin{array}{llll}
M_{B_d}=5.28{\rm GeV},& m_{b}=4.66{\rm GeV},&M_{K^{*\pm}}=0.89{\rm
GeV}, &\tau_{B_d^0}=1.536{\rm ps},\\f_{B_d}=200\pm30{\rm MeV},
&f_{K}=160{\rm MeV},&f_{K^*}=218\pm4{\rm MeV},&f^\bot_{K^*}=175\pm25{\rm MeV},\\
V_{ud}=1-\lambda^2/2,&V_{ub}=A\lambda^3(\rho-i\eta),
&V_{td}=A\lambda^3(1-\rho-i\eta),&V_{tb}=1.
\end{array}
\end{equation}
The Wolfenstein parameters of the CKM matrix elements are taken
as~\cite{s.eidelman}:
$A=0.8533\pm0.0512,~\lambda=0.2200\pm0.0026,~\bar\rho=0.20\pm0.09,
~\bar\eta=0.33\pm0.05$, with
$\bar\rho=\rho(1-\frac{\lambda^2}{2}),~\bar\eta=\eta(1-
\frac{\lambda^2}{2})$. The coefficient of the twist-3 distribution
amplitude of the pseudoscalar $K$ meson is chosen as
$\mu_K=\mu_\pi =1.5\pm0.2$~GeV~\cite{M,bn}.

\section*{Appendix B: The decay amplitudes}

To evaluate the hadronic matrix elements, we may adopt the QCD
factorization formalism. For the annihilation process $\bar B \to
M_1M_2$, the matrix element can be written as~\cite{bn}
\begin{equation}
\langle M_1M_2|O_i|\bar{B}^0\rangle=f_B\Phi_B\ast
f_{M_1}\Phi_{M_1}\ast f_{M_2} \Phi_{M_2}*T_i,\label{qcdm}
\end{equation}
where $O_i$ is the effective operator appearing in the effective
weak Hamiltonian, the $\ast$ products imply  integrations over the
light-cone momentum fractions of the constituent quarks inside the
relevant mesons. $T_i$ is the hard-scattering kernel that can be
computed perturbatively with the QCD factorization approach.
$\Phi_M$ and $f_i$ are the leading-twist light-cone distribution
amplitudes and the decay constants, respectively.

Using the formula (\ref{qcdm}) and the meson wave functions given
in~\cite{h.y.cheng,M.T,Genon,p.ball}, we can evaluate the decay
amplitudes of the $W$-exchanged diagrams in Fig.~\ref{tree}~(only
the $O_1$ operator has contribution) and the penguin annihilation
diagrams in Fig.~\ref{penguin}~($O_4$, $O_6$, $O_8$, and $O_{10}$
have contributions).

For final states containing two pseudoscalar $K$ meson, the
amplitudes are found to be
\begin{eqnarray}
{\cal A}_T(\bar{B}_{d}\to K^{+}K^{-})&=&C_1\times A_1,
\nonumber \\
{\cal A}_P(\bar{B}_{d}\to K^{+}K^{-})&=&\biggl(2C_4
+\frac{C_{10}}{2}\biggl)\times A_1 +\biggl(2C_6
+\frac{C_{8}}{2}\biggl)\times A_2,\label{amplitude}
\end{eqnarray}
where
\begin{eqnarray}
A_1&=&\frac{G_{F}}
{\sqrt{2}}f_{B}f_K^2\pi\alpha_s(\mu)\frac{C_F}{N_C^2}
\int_0^{1}d\xi\int_0^1dx\int_0^1dy\Phi_1^B(\xi) \nonumber\\& \times
& \biggl\{\biggl(x\frac{M_B^4}{D_dk^2}
+(y+\xi-\xi_B)\frac{M_B^4}{D_bk^2}
\biggl)\Phi_K(x)\Phi_K(y)\nonumber\\ &+&\frac{\mu_K^2}{M^2_B}
\biggl[\biggl((x+y-\xi)\phi_k(x)\phi_k(y)+(y-\xi-x)\phi_k(x)
\frac{\phi'_\sigma(y)}{6}\nonumber\\&
-&(y-\xi-x)\phi_k(y)\frac{\phi'_\sigma(x)}{6}-(x+y-\xi)
\frac{\phi'_\sigma(x)\phi'_
\sigma(y)}{36}\biggl)\frac{M_B^4}{D_dk^2}\nonumber\\ &
+&\biggl((x+y+\xi-2\xi_B)\phi_k(x)\phi_k(y)+(y+\xi-x)\phi_k(x)
\frac{\phi'_\sigma(y)}{6}\nonumber\\
&-&(y+\xi-x)\phi_k(y)\frac{\phi'_\sigma(x)}{6}+(x+y+\xi-2)\frac{\phi'_
\sigma(x)\phi'_\sigma(y)}{36}\biggl) \frac{M_B^4}{D_bk^2}
\biggl]\biggl\},\label{pp1}
\end{eqnarray}
\begin{eqnarray}
A_2&=&\frac{G_{F}}{\sqrt{2}}
f_{B}f_K^2\pi\alpha_s(\mu)\frac{C_F}{N_C^2}
\int_0^{1}d\xi\int_0^1dx\int_0^1dy\Phi_1^B(\xi)\nonumber\\
&\times&\biggl\{\biggl((y-\xi) \frac{M_B^4}{D_dk^2}
+(x-\xi_B)\frac{M_B^4}{D_bk^2}
\biggl)\Phi_K(x)\Phi_K(y)\nonumber\\
&+&\frac{\mu_K^2}{M^2_B}\biggl[\biggl((x+y-\xi)\phi_k(x)
\phi_k(y)-(y-\xi-x)\phi_k(x)\frac{\phi'_\sigma(y)}{6}\nonumber\\&
+&(y-\xi-x)\phi_k(y)\frac{\phi'_\sigma(x)}{6}-(x+y-\xi)
\frac{\phi'_\sigma(x)\phi'_
\sigma(y)}{36}\biggl)\frac{M_B^4}{D_dk^2}\nonumber\\ &
+&\biggl((x+y+\xi-2\xi_B)\phi_k(x)
\phi_k(y)-(y+\xi-x)\phi_k(x)\frac{\phi'_\sigma(y)}{6}
\nonumber\\&+&(y+\xi-x)\phi_k(y)\frac{\phi'_\sigma(x)}{6}+(x+y+\xi-2)
\frac{\phi'_ \sigma(x)\phi'_\sigma(y)}{36}\biggl)
\frac{M_B^4}{D_bk^2} \biggl]\biggl\},\label{pp2}
\end{eqnarray}
where $\xi_B=(M_B-m_b)/M_B$ with $M_B$ being the mass of $B_d$
meson, $\phi^{'}_\sigma(x)=\frac{d\phi_\sigma(x)}{dx}$, $\Phi's$ and
$\phi's $ are the leading twist~(twist-two) and twist-three
light-cone distribution amplitudes of mesons, respectively. We set
the scale $\mu$ to be the averaged virtuality of the time-like
gluon, i.e., $\mu=m_{b}/2$. $k^2$ and $D_{b,d}$ arise from the
propagators of the virtual gluon, the bottom quark $b$, and the down
quark $d$, respectively.

For $\bar B_d\to K^\pm K^{*\mp}$ decays, we need only consider the
longitudinal wave function of the vector $K^*$ meson, due to the
conservation of the angular momentum. The corresponding decay
amplitudes of $\bar B_d\to K^+K^{*-}$ are given with the same form
as Eq.~(\ref{amplitude}), but with the amplitudes of $A_1$ and
$A_2$ replaced by
\begin{eqnarray}
A_1&=&\frac{G_{F}}
{\sqrt{2}}f_{B}f_K\pi\alpha_s(\mu)\frac{C_F}{N_C^2}
\int_0^{1}d\xi\int_0^1dx\int_0^1dy\Phi_1^B(\xi) \nonumber\\& \times
& \biggl[f_{K^*}\biggl(x\frac{M_B^4}{D_dk^2}
+(y+\xi-\xi_B)\frac{M_B^4}{D_bk^2}\biggl)\Phi_K(x)\Phi_K(y)\nonumber\\
&+&f_{K^*}^\bot\mu_K\frac{m_{K^*}}{M_B}\biggl((-x+y-\xi)\frac{h^{'(s)
}_\|(y)\phi'_\sigma(x)}{12}-(x+y-\xi)\frac{h^{'(s)}_\|(y)\phi(x)}{2}
\nonumber\\&+&(x+y-\xi)\frac{\phi'_\sigma(x)h^{(t)}_\|(y)}{3}+2(x-y+\xi)
\phi(x)h^{(t)}_\|(y)\biggl)\frac{M_B^4}{D_dk^2}\nonumber\\
&+&f_{K^*}^\bot\mu_K\frac{m_{K^*}}{M_B}\biggl((-x+y-\xi)\frac{h^{'(s)
}_\|(y)\phi'_\sigma(x)}{12}-(x+y+\xi-2\xi_B)\frac{h^{'(s)}_\|(y)\phi(x)}{2}
\nonumber\\&-&(x+y+\xi-2)\frac{\phi'_\sigma(x)h^{(t)}_\|(y)}{3}+(x-y-\xi)
\phi(x)h^{(t)}_\|(y)\biggl)\frac{M_B^4}{D_bk^2}
 \biggl],\label{pv1}
\end{eqnarray}

\begin{eqnarray}
A_2&=&\frac{G_{F}}
{\sqrt{2}}f_{B}f_K\pi\alpha_s(\mu)\frac{C_F}{N_C^2}\int_0^{1}d\xi
\int_0^1dx\int_0^1dy\Phi_1^B(\xi)\nonumber\\& \times &
\biggl[f_{K^*}\biggl((-y+\xi)\frac{M_B^4}{D_dk^2}
+(x+\xi_B)\frac{M_B^4}{D_bk^2}\biggl)\Phi_K(x)\Phi_K(y)\nonumber\\
&+&f_{K^*}^\bot\mu_K\frac{m_{K^*}}{M_B}\biggl((-x+y-\xi)\frac{h^{'(s)
}_\|(y)\phi'_\sigma(x)}{12}+(x+y-\xi)\frac{h^{'(s)}_\|(y)\phi(x)}{2}
\nonumber\\&-&(x+y-\xi)\frac{\phi'_\sigma(x)h^{(t)}_\|(y)}{3}+2(x-y+\xi)
\phi(x)h^{(t)}_\|(y)\biggl)\frac{M_B^4}{D_dk^2}\nonumber\\
&+&f_{K^*}^\bot\mu_K\frac{m_{K^*}}{M_B}\biggl((-x+y-\xi)\frac{h^{'(s)
}_\|(y)\phi'_\sigma(x)}{12}+(x+y+\xi-2\xi_B)\frac{h^{'(s)}_\|(y)\phi(x)}{2}
\nonumber\\&+&(x+y+\xi-2)\frac{\phi'_\sigma(x)h^{(t)}_\|(y)}{3}+(x-y-\xi)
\phi(x)h^{(t)}_\|(y)\biggl)\frac{M_B^4}{D_bk^2}
 \biggl],\label{pv2}
\end{eqnarray}
where $h^{'(s)}_{\parallel}(y)=\frac{dh^{(s)}_{\parallel}(y)}{dy}$.
With the change for the signs of the second and the third terms in
the twist-3 amplitudes in Eqs.~(\ref{pv1}) and~(\ref{pv2}) and the
exchange of the variable $x$ and $y$ of the wave functions, we can
get the corresponding decay amplitudes of $\bar B_d\to K^-K^{*+}$
mode.

Finally, for the $\bar B\to K^{*+}K^{*-}$ decay, in the rest frame
of $\bar{B}^0$ system, we have $\lambda_1=\lambda_2=\lambda$~(where
$\lambda_1$ and $\lambda_2$ denote the helicities of the $K^{*+}$
and $K^{*-}$ mesons, respectively) through the helicity conservation
since the $\bar{B}^0$ meson has helicity zero. So, there are
generally three decay amplitudes, $H_0$, $H_+$, and $H_-$,
representing $\lambda=0$, $1$, and $-1$, respectively. Considering
only the leading twist contributions, we can get the longitudinal
amplitudes ${\cal A}_{0T}$ and ${\cal A}_{0P}$ of the decay $\bar
B_d\to K^{*+}K^{*-}$ from Eqs.~(\ref{pp1}) and~(\ref{pp2}) by
keeping only the $\Phi_K(x)\Phi_K(y)$ term and replacing the decay
constant $f_K$ by $f_{K^*}$. The total longitudinal amplitude is
then given as
\begin{equation}
H_{0}=V_{ub}V_{ud}^*{\cal A}_{0T}-V_{tb}V_{td}^*{\cal A}_{0P}.
\end{equation}
As for the transverse amplitude $H_{\pm}$, we have
\begin{eqnarray}
H_{+}&=&\frac{G_{F}} {\sqrt{2}}f_{B}f_{K^*}^\bot
\frac{m_{K^*}^2}{M_B^2}\pi\alpha_s(\mu)\frac{C_F}{N_C^2}
\int_0^{1}d\xi\int_0^1dx\int_0^1dy\Phi_1^B(\xi)\biggl(C_1+
2C_4+\frac{C_{10}}{2}\biggl)\nonumber\\& \times&
\biggl((-f(x)g^{(v)}_\bot(y)
-g^{(v)}_\bot(x)g^{(a)}_\bot(y)/4+f(x)g^{'(a)}_\bot(y)/8
+g^{'(a)}_\bot(x)g^{(a)}_\bot(y)/32)\frac{M_B^4}{D_dk^2}\nonumber\\
&+& (-f(x)g^{(v)}_\bot(y)
+g^{(v)}_\bot(x)g^{(a)}_\bot(y)/4-f(x)g^{'(a)}_\bot(y)/8
+g^{'(a)}_\bot(x)g^{(a)}_\bot(y)/32)\frac{M_B^4}{D_bk^2}\biggl)\nonumber\\
&+&\frac{G_{F}} {\sqrt{2}}f_{B}f_{K^*}^\bot
\frac{m_{K^*}^2}{M_B^2}\pi\alpha_s(\mu)\frac{C_F}{N_C^2}
\int_0^{1}d\xi\int_0^1dx\int_0^1dy\Phi_1^B(\xi)\biggl(
2C_6+\frac{C_{8}}{2}\biggl)\nonumber\\& \times&
\biggl((-g^{(v)}_\bot(x)f(y)
+g^{(a)}_\bot(x)g^{(v)}_\bot(y)/4-g^{'(a)}_\bot(x)f(y)/8
+g^{(a)}_\bot(x)g^{'(a)}_\bot(y)/32)\frac{M_B^4}{D_dk^2}\nonumber\\
&+& (-g^{(v)}_\bot(x)f(y)
-g^{(a)}_\bot(x)g^{(v)}_\bot(y)/4+g^{'(a)}_\bot(x)f(y)/8
+g^{(a)}_\bot(x)g^{'(a)}_\bot(y)/32)\frac{M_B^4}{D_bk^2}\biggl),\label{vv}
\end{eqnarray}
where the function
$f(x)=\int_0^xdu\left(\phi_\|(u)-g^{(v)}_\bot(u)\right)$ and
$g^{'(a)}_\perp(y)=\frac{dg^{(a)}_\perp(y)}{dy}$. Changing the signs
of the $g^{(a)}_\bot g^{(v)}_\bot$ and $g^{'(a)}_\bot f$ terms in
Eq.~(\ref{vv}), we can get the other amplitude $H_{-}$.

\end{document}